\documentclass[english,12pt]{article}
\usepackage{array}
\usepackage{pstricks,pst-node,pst-text,pst-3d}
\usepackage{delarray}
\usepackage{t1enc}
\usepackage[utf8]{inputenc}
\usepackage[english]{babel}
\usepackage{graphicx,multirow}
\usepackage{eepic}
\usepackage{epic}
\usepackage{euscript}
\usepackage{indentfirst}
\usepackage{amsmath}
\usepackage{a4wide}
\usepackage{ulem}
\usepackage{url}
\usepackage{amsfonts}
\usepackage{amssymb}
\usepackage{stmaryrd}
\usepackage{theorem}
\usepackage{verbatim}
\usepackage{multirow}
\usepackage[numbers,sort&compress]{natbib}
\usepackage{tabularx}
\usepackage{caption}
\usepackage{subcaption}
\usepackage{float}
\usepackage{pgfplots}

\unitlength=1mm{}

\begin{document}

\title{Forecasts in Schelling's segregation model\thanks{I thank Pablo Jensen and Julien Flaig for their valuable remarks.}}

\author{
Nicolas \textsc{Houy}\thanks{University of Lyon, Lyon, F-69007, France;
CNRS, GATE Lyon Saint-Etienne, F-69130, France. Email: \texttt{houy@gate.cnrs.fr}.}
}

\date{\today}

\maketitle

\begin{abstract}
In Schelling's segregation model \cite{Schelling1971}, the successive moves of agents optimizing their own locations lead to a suboptimal segregated distribution of the population, even though all agents have the same preference for mixed neighborhoods. One of the main assumptions underlying this general result of segregation models is that agents rely on comparisons between instantaneous utilities in order to make their moving decisions. On the contrary and certainly more reasonably, we assume in this article that agents forecast later states using a linear extrapolation of past states heuristic in order to make their decisions. We show that for a relatively small set of parameters, considering forecasting agents allows to dramatically reduce sub-optimality in a framework close to \cite{Schelling1971}'s model.
\end{abstract}

\medskip 

Declarations of interest: none.

\pagebreak
\baselineskip=6mm
\section{Introduction}

We now know from many experiments as well as many theoretical works that agents making decisions in a decentralized way can lead to a socially suboptimal situation. Externalities are one important reason that can explain why micro motives can lead to bad macro behavior \cite{SchellingBook}. Externalities are present when the actions of an individual may have an impact on another individual's welfare. Then, the centralized optimization problem, taking into account this impact can hardly have the same results as the combination of decentralized optimization problems not taking into account the externalities, by definition of what an externality is.

One particular example of this case is the segregation model proposed by \cite{Schelling1971}, using a framework close to \cite{Sakoda1971} and on which \cite{Grauwin2009}, \cite{Jensen2018} and \cite{Flaig2019} elaborated recently. We will use the formulation introduced in the two latter studies for the sake of its simplicity. In this agent-based model, all agents have the same preferences for living in a mixed city, \textit{i.e.} in neighborhoods with intermediate density. However, the dynamics of decentralized decision making individuals generally leads to a collection of overpopulated and empty neighborhoods. Then, even though all individuals have the same, therefore non conflicting, social and individual goals, the social optimum is not attained by the society by means of decentralized decision making. Economically, the reason is that when an agent changes its location, she does not take into account the resulting situation for the individuals living in the neighborhood she is leaving. This is where the externality lies. Even though the global goal is well-defined and the same for all individuals, her own welfare may locally contradict the social welfare. When making a choice, she does not simply compute her share of the global optimization problem as she would if there was no externality. Notice that, even though it is in a radically different dynamic setting, it is the same reasoning that explains why the absence of externality is a necessary condition for the first fundamental theorem of welfare economics to apply (see \cite{mas-colell95}).

Another reason for which the segregation models cited above miss the social common target and lead to suboptimal situation is that individuals have some sort of bounded rationality. Indeed, in all these models, agents make their decisions based on the current situation whereas the situation that is used to measure the final welfare --~the one that is suboptimal and that should somehow be only an aggregation of the agents' utility functions~-- is the one obtained at stationary state, once an equilibrium is reached. For an agent to be fully rational in this setting, she should anticipate the future possible states and integrate these expectations in her computation for decision-making. In most studies directly related to \cite{Schelling1971}, when an agent has an opportunity to move, she compares the current utility she would get from moving to the current utility she would get from staying. That is fine if we consider either that agents have infinite discount factor or if they are myopic to the future. If both of this assumptions are not accepted, they should try to anticipate what will come next and evaluate their intertemporal utilities from moving or staying.

To the best of our knowledge, only \cite{BenitoOstolaza2015} consider fully rational agents in a particular segregation model setting. However, their results can hardly be generalized beyond the precise model they use with a ring city. Indeed, considering decisions of agents able to make expectations in a model is computationally complex. Indeed, on the one hand, expectations influence choices and hence the dynamics of the system. But in the other hand, the dynamics should somehow influence the expectations for acceptably rational agents. In simple continuous cases, the theory of rational expectations is used by economists (\cite{sargent1987rational}) in order to technically transform this problem in the analytical search for a (functional) fixed point. In some more complex systems with continuous dynamics, special solving algorithms are introduced (\cite{Flaig2018} for one example). In the current study, we will rely on an extrapolation heuristics. Agents forecasts future states over some time horizon by linearly extrapolating from recent history. This way, forecasts are made looking backward and the complexity to have both expectations and dynamics coupled vanishes and our complex problem can be solved with stochastic simulations. We do not lose the full richness of the segregation models.

As a consequence of considering expectations as an extrapolation of past states, we will be able to show that for some small domains of parameters, the inefficiencies of the \cite{Schelling1971} model can be dramatically reduced. This result sets a bound for the main result of segregation models --~sub-optimality of the sequence of individual decisions~--in a direction --~forecasting agents~-- that has been overlooked until now.

In Section \ref{sec:model}, we present the model extensively borrowed from \cite{Jensen2018} with the formal description of how we expect agents to form their expectations. In Section \ref{sec:results}, we display our results and concentrate on a case where segregation is almost completely non present at stationary state. Finally, Section \ref{sec:conclu} concludes.

\section{Materials and methods}\label{sec:model}

\paragraph{Dynamics.} Our formal framework is extensively borrowed from \cite{Jensen2018}. We consider a set of agents moving from one neighborhood to another in a city. The city is divided in $Q$ neighborhoods ($Q=36$ throughout the present article). Each neighborhood can accommodate $H$ agents ($H=225$ throughout the present article) in housing sites. The number of agents is $N$ ($N=H \times Q \times 0.4=3,240$ throughout the present article). The initial state is stochastic and generated by randomly drawing with uniform probability an initial neighborhood for each agent (satisfying the constraint that the number of agents in each neighborhood be smaller than $H$).\footnote{In Appendix, we display results for different initial conditions. Results remain qualitatively valid even though base value are different.} As time continuously passes by, each agent is drawn by chance following a Poisson Process with rate $\lambda$. $\lambda$ is normalized to be 1. Once an agent is considered, she is matched with an empty housing site uniformly picked at random. The agent can then decide to move to the designated house using the decision rule extensively described below.

In any given neighborhood, an agent has an instantaneous utility $u(\rho)$ that depends on density $\rho$.\footnote{The density of a neighborhood is the number of individuals living in this neighborhood divided by the number of accommodation sites in this neighborhood, $H$.} She gets a maximum utility for $\rho=1/2$ and then loses utility with constant marginal effect as density gets further away from $1/2$. Formally, after normalization so that $u(0)=u(1)=0$ and $u(1/2)=1$:
$$\forall \rho \in [0,1], u(\rho)=\left\{\begin{array}{ll}2.\rho & \text{ if }\rho<0.5, \\ 2.(1-\rho) & \text{ otherwise.}\end{array}\right.$$

\paragraph{Moving decisions.} Let us consider an agent from neighborhood $i$ who has a chance to move to neighborhood $j$ at time $T$. First, the agent forecasts the future population densities in each neighborhood by applying a linear extrapolation using the following two points: the current population density and the one at time $T-\underline{m}$.\footnote{Notice that when considering the current population in $j$, the agent considers her move in this neighborhood.} We will call parameter $\underline{m}$ the memory length. Formally, if $\rho_i^T$ and $\rho_i^{T-\underline{m}}$ are the population densities observed in neighborhood $i$ at times $T$ and $T-\underline{m}$ respectively, the population density at time $t>T$ forecast by an agent is given by\footnote{For any real number $r$, $\left[r\right]_0^1=\max(0,\min(r,1))$ is the identity function with bounds in 0 and 1. Also, if $T-\underline{m}<0$, we consider the past population in 0.}
$$\rho^*_i(t)=\left[ \rho_i^T+\frac{\rho_i^T-\rho_i^{T-\underline{m}}}{\underline{m}}.(t-T) \right]_0^1.$$
Now, at time $t$, the agent expects total future utility $U^*_i(t)$ to be the discounted sum of the instantaneous utility applied to her population density forecast over a given horizon $\overline{m}$ that we will call the forecast length. Formally,
$$U^*_i(t)=\int_t^{t+\overline{m}} e^{-\beta.(\theta-t)}u(\rho^*_i(\theta))d\theta$$
where $\beta$ is the discount factor. Throughout this article, we will consider $\beta=0.03$.\footnote{In a standard general equilibrium model, $\beta$ should be close to the economy interest rate.} Finally, the agent accepts to move from neighborhood $i$ to neighborhood $j$ at time $t$ if and only if $U^*_j(t)>U^*_i(t)$.

Notice that the heuristics we consider can be labeled extrapolating in the sense that expectations are computed using past observations only. Notice that the agents we consider are not fully rational in the sense usually used in economics. In particular, they are not rational in the sense of rational expectations that forecast the states that will actually occur in the future, see \cite{sargent1987rational, mas-colell95}.

\section{Results}\label{sec:results}

\begin{figure}[ht]
\centering
\begin{subfigure}{.45\textwidth}
  \centering
  \includegraphics[width=.9\linewidth]{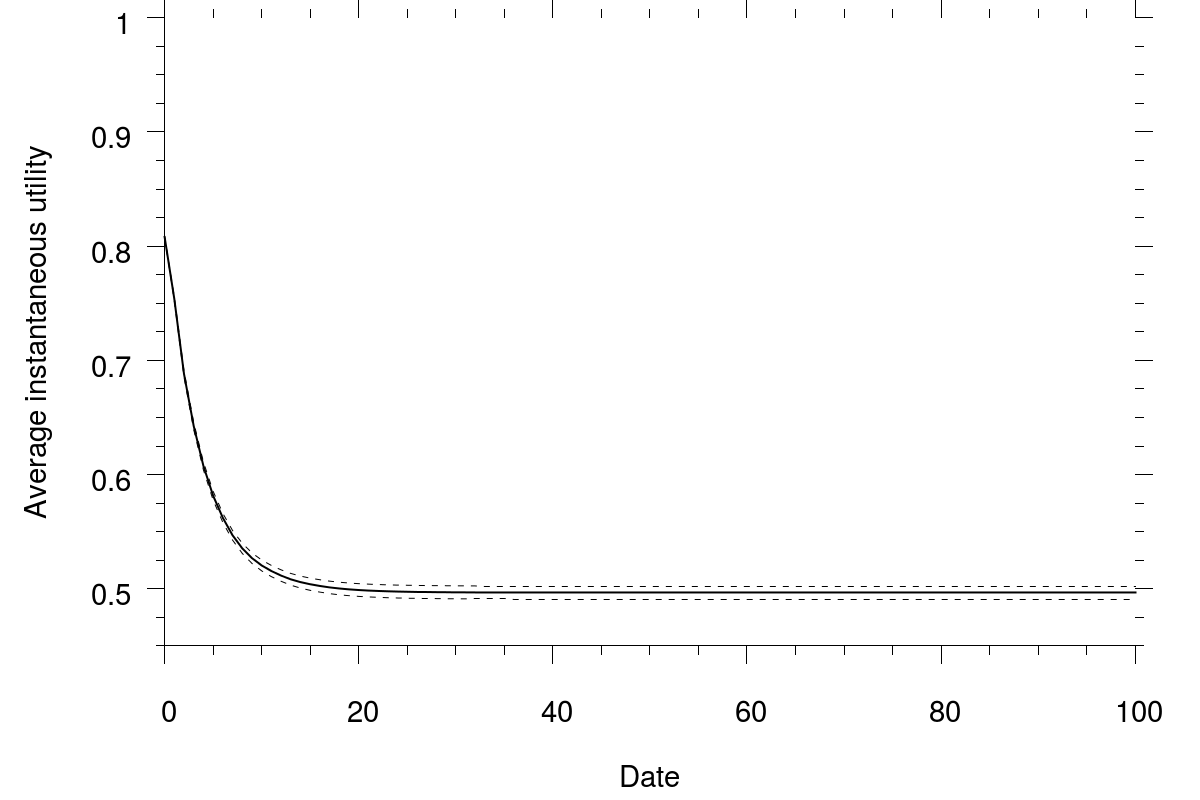}
\end{subfigure}%
\begin{subfigure}{.45\textwidth}
  \centering
  \includegraphics[width=.9\linewidth]{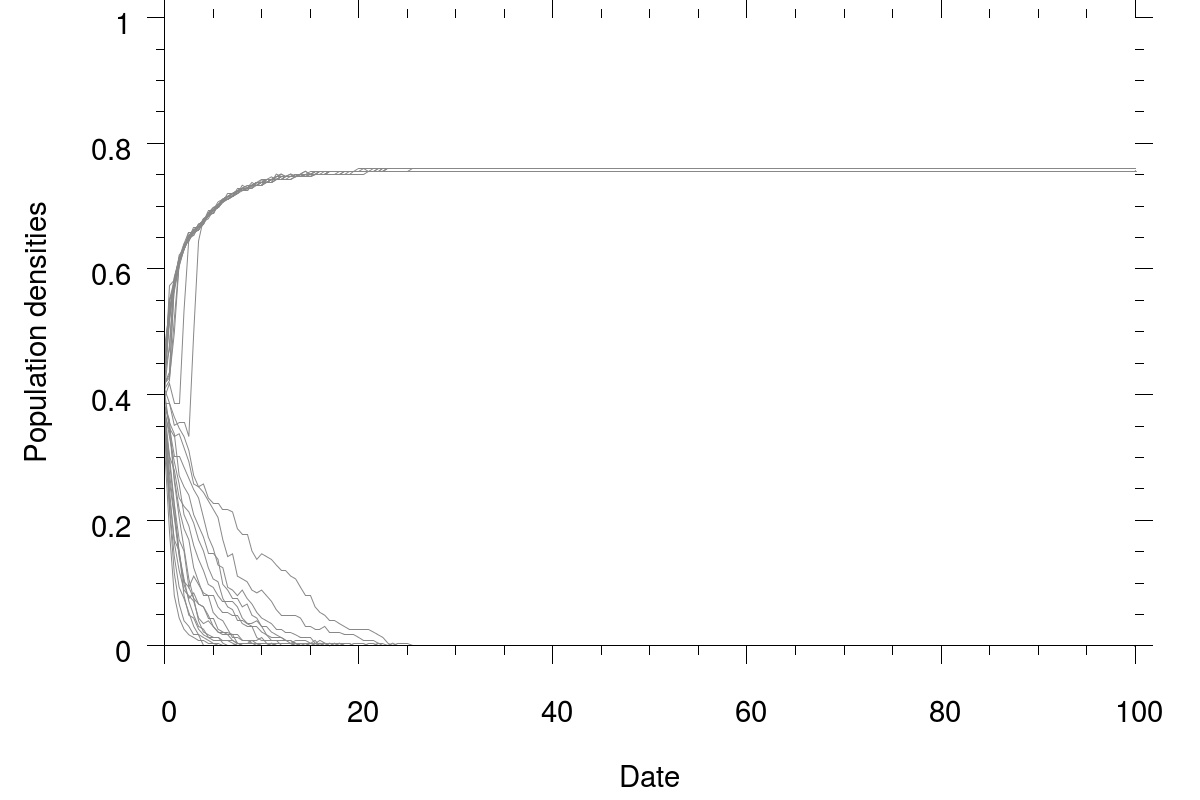}
\end{subfigure}
\caption{Model dynamics with $\underline{m}=10^{-2}$ and $\overline{m}=10^{-4}$. \textit{Left:} Average instantaneous utility with 95~\% confidence interval computed over 120 repetitions. \textit{Right:} Neighborhoods population densities for one repetition.}
\label{fig1}
\end{figure}

Let us first consider the extreme case with the forecast length $\overline{m}$ arbitrarily small and the memory length $\underline{m}$ not too small. In this case, comparisons between neighborhoods are made as if depending on the instantaneous utility obtained in each neighborhood and we find the same results, displayed in Figure \ref{fig1}, as in \cite{Jensen2018}. At the start, we have population densities close to 40~\% in all neighborhoods and hence an average instantaneous utility close to 0.8. However, even though all individuals have the same objective to live in a neighborhood with population density 50~\%, the population stabilizes with empty neighborhoods and neighborhoods with densities between 75~\% and 80~\%. Hence, a stationary state is reached with average instantaneous utility of about 0.5.

Now, let us consider the extreme case with the memory length $\underline{m}$ arbitrarily small and the forecast length $\overline{m}$ not too small. In this extreme case, any observed change in population density for a neighborhood leads to a short-term forecast of 0 or 1 for future population density in this neighborhood and hence a small intertemporal utility. Hence, no move --~implying a change in the destination neighborhood population density~-- is ever implemented, the population densities are constant at about 40~\% and the average instantaneous utility is close to 0.8.

\begin{figure}[ht]
\centering
\includegraphics[width=.9\linewidth]{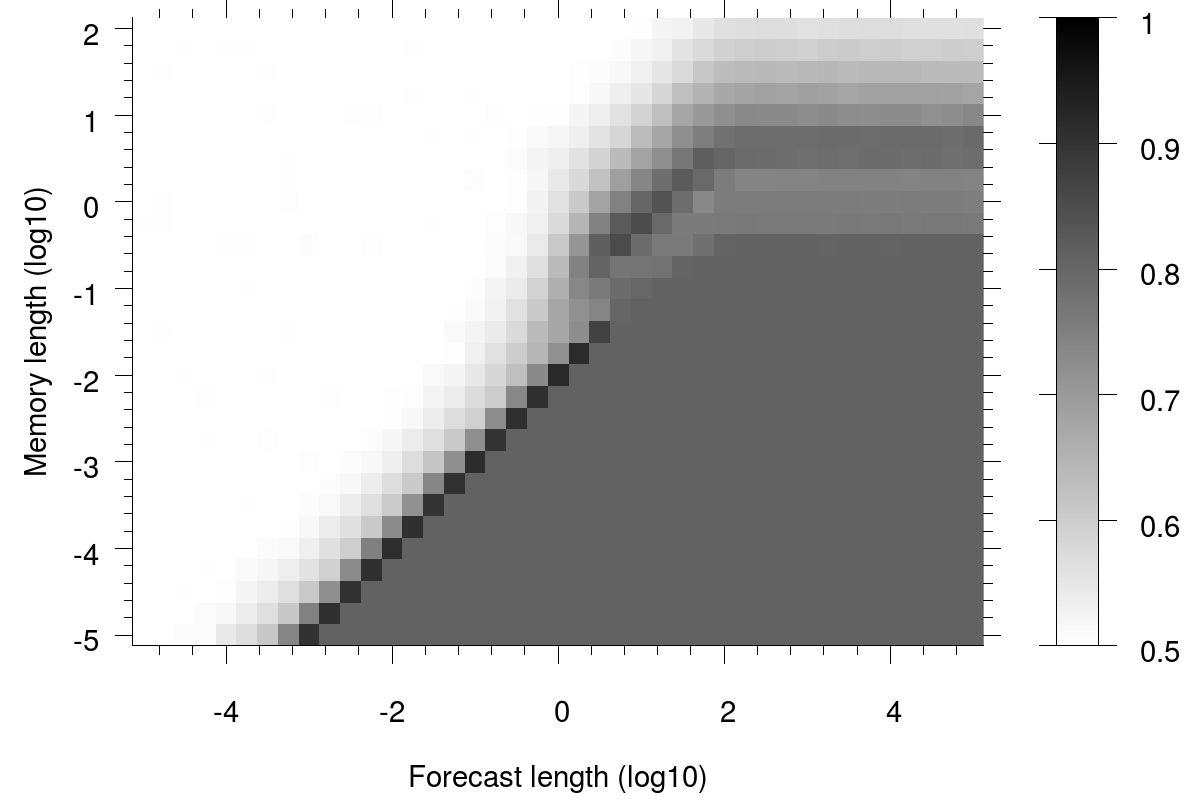}
\caption{Average instantaneous utility at stationary value as a function of the memory length $\underline{m}$ and the forecast length $\overline{m}$. Values averaged over 120 repetitions.}
\label{fig2}
\end{figure}

In Figure \ref{fig2}, we display the average instantaneous utility at its stationary value as a function of the memory length $\underline{m}$ and the forecast length $\overline{m}$.

\begin{figure}[ht]
\centering
\begin{subfigure}{.45\textwidth}
  \centering
  \includegraphics[width=.9\linewidth]{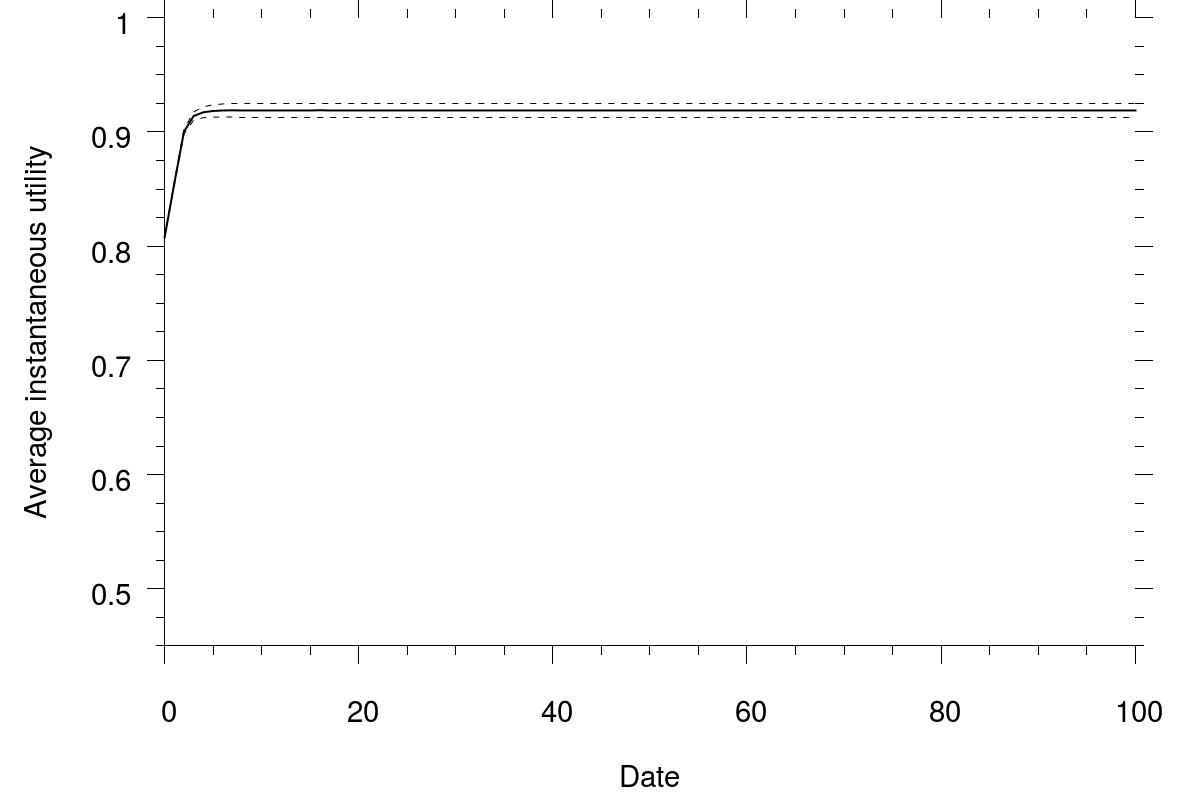}
\end{subfigure}%
\begin{subfigure}{.45\textwidth}
  \centering
  \includegraphics[width=.9\linewidth]{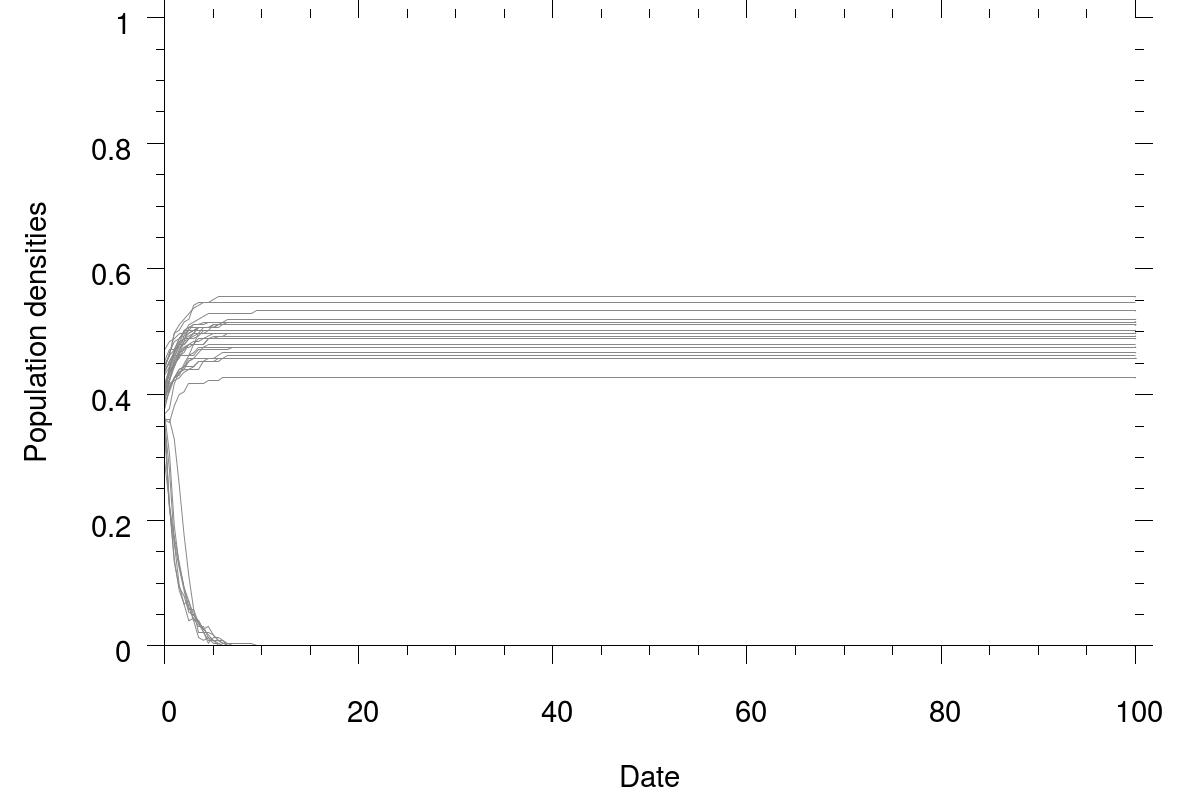}
\end{subfigure}
\caption{Model dynamics with $\underline{m}=10^{-2}$ and $\overline{m}=1$. \textit{Left:} Average instantaneous utility with 95~\% confidence interval computed over 120 repetitions. \textit{Right:} Neighborhoods population densities for one repetition.}
\label{fig3}
\end{figure}

We can see that the transition from the extreme cases noted above is not monotonic and there are cases with average instantaneous utility larger than the 0.8 mark. Hence, memory and expectations can solve the problem of externalities that makes a society of individuals with the same personal objective (living in a 50~\% density neighborhood) socially miss this objective. In the following, we consider the case with memory length $\underline{m}=10^{-2}$ and forecast length $\overline{m}=1$. In this case, at stationary state, the average instantaneous utility reaches 0.92 (95~\% CI: 0.91~--~0.93). In Figure \ref{fig3}, we display the average instantaneous utility and neighborhoods population densities in this case.

When $\underline{m}=10^{-2}$ and $\overline{m}=1$, the population spreads very quickly between a number of neighborhoods with population density close to 50~\%, hence reaching a situation with average instantaneous utility close to 1. Only relatively small variations around density 50~\% prevents the situation to be optimal.

\begin{figure}[ht]
\centering
\begin{subfigure}{.45\textwidth}
  \centering
  \includegraphics[width=.9\linewidth]{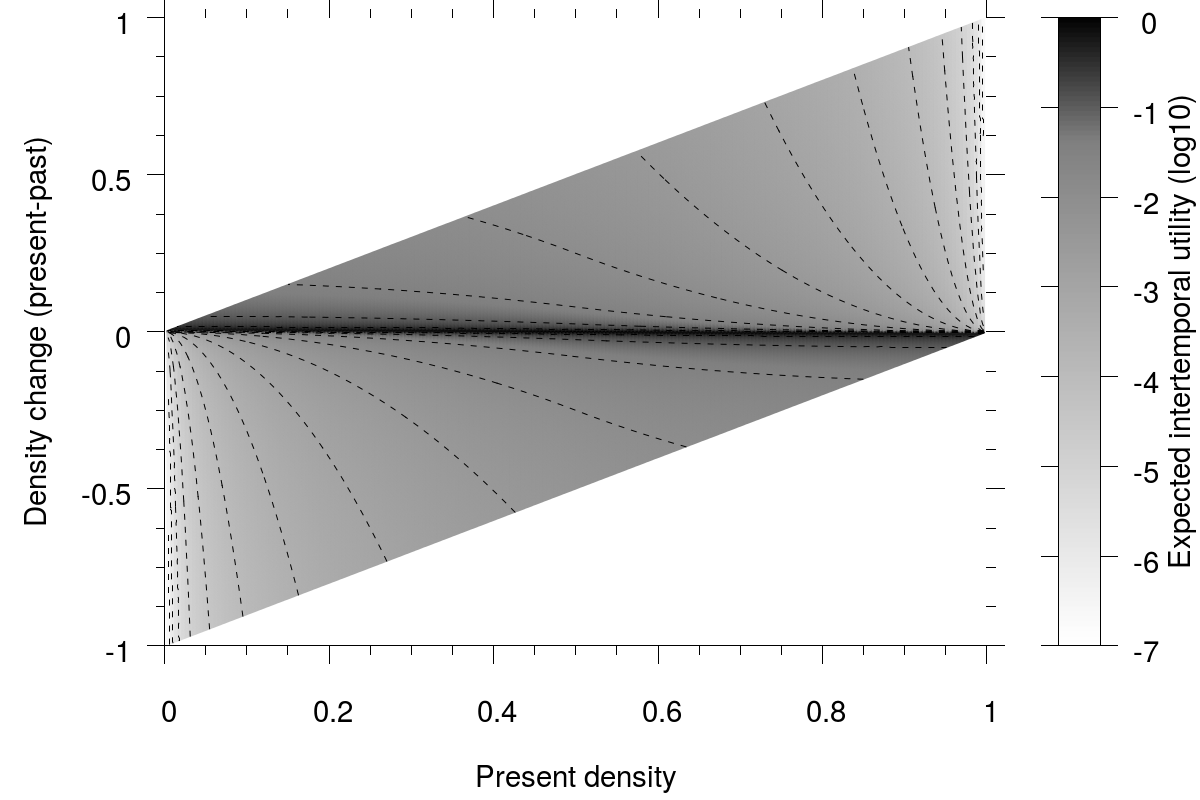}
\end{subfigure}%
\begin{subfigure}{.45\textwidth}
  \centering
  \includegraphics[width=.9\linewidth]{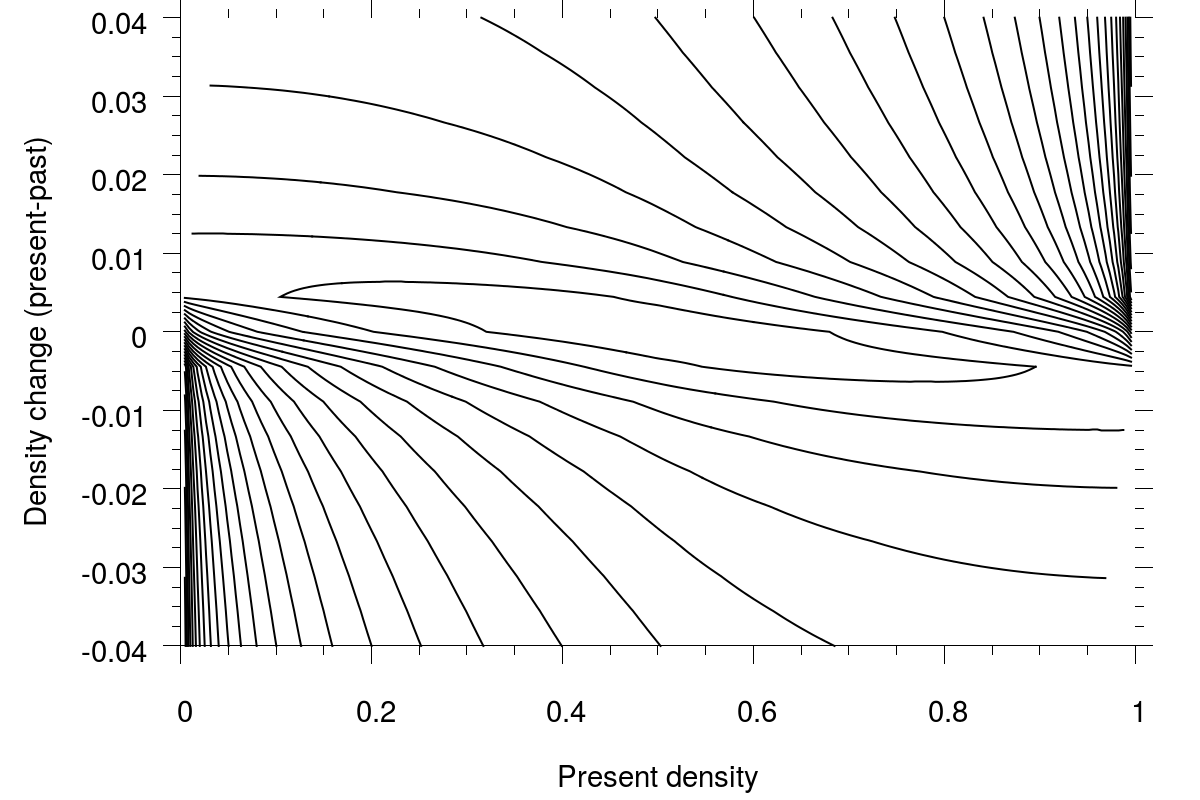}
\end{subfigure}
\caption{\textit{Left:} Expected intertemporal (computed over $\overline{m}=1$) utility from living in a neighborhood as a function of the present population density in this neighborhood and the change in population density in this same neighborhood over the past $\underline{m}=10^{-2}$. Dashed lines for identical values contours. \textit{Right:} Same zoomed around 0 ordinate.}
\label{fig4}
\end{figure}

In order to go further in our reasoning, let us describe Figure \ref{fig4}. Given $\underline{m}$ and $\overline{m}$, the intertemporal expected utility an agent computes at any time $t$ depends only on the population density in the past (at $t-\underline{m}$) and at time $t$. In Figure \ref{fig4}, we display the expected (with $\overline{m}=1$) intertemporal utility as a function of both the present population density in a neighborhood and the density in the same neighborhood $\underline{m}$ before, \textit{i.e.} in the memory.

Now, let us explain Figure \ref{fig4} that is not obvious to interpret. Let us consider a given neighborhood with 50~\% current population density and 50~\% past population density. Since no variation in the past is observed on average, agents expect that the density will remain unchanged in the $\overline{m}$ long future. Hence, the expected intertemporal utility is maximal with a constant 1 instantaneous utility. Now, if the observed past population is greater than 50~\% (negative change), a future decrease in population density is expected and hence a decreased expected intertemporal utility. The same occurs for an observed past population smaller than 50~\%. Now, let us consider a current population density of 40~\%. A past negative change of population density implies an expected population density decreasing even further in the future and hence a fast decreasing expected intertemporal utility. On the contrary, a small past positive change of population density implies an expected future population density that will get close to 50~\% and hence a high expected intertemporal utility. Of course, if this positive change is too large, it is also expected that the population density will increase until it reaches values that bring very low instantaneous utility.

Now, let us try and give some hints to explain the dynamics displayed in Figure \ref{fig3} with the following steps.
\begin{enumerate}
\item \label{st1} In the beginning, individuals consider the possibility to move from a neighborhood with a population density constant and close to 40~\% to a neighborhood with a population density close to 40~\% and increasing due to the extrapolation of their own move. Given the parameters, they do accept to move.
\item \label{st2} However, as population tends to concentrate in a few neighborhoods, the ones growing fastest and with density closest to 50~\% are the less attractive ones since it is expected that they will bring lower future streams of utility.
\item \label{st3} As population settles and memory integrates the fact that there was no recent move, decisions not to move are unchanged.
\end{enumerate}

In the extreme case with $\underline{m}$ arbitrarily small and the forecast length $\overline{m}$ not too small, incentives described in Step \ref{st1} are not enough for a move to be considered anything but going in a neighborhood that will be overcrowded pretty soon. Hence, no move ever takes place. In the extreme case with the forecast length $\overline{m}$ arbitrarily small and the memory length $\underline{m}$ not too small, moving into neighborhoods with high population densities is made more interesting than as described in Step \ref{st2} by removing the forecast part of the reasoning. Hence, the overcrowded neighborhoods in the stationary states in this case.

\begin{figure}[ht]
\centering
\begin{subfigure}{.45\textwidth}
  \centering
  \includegraphics[width=.9\linewidth]{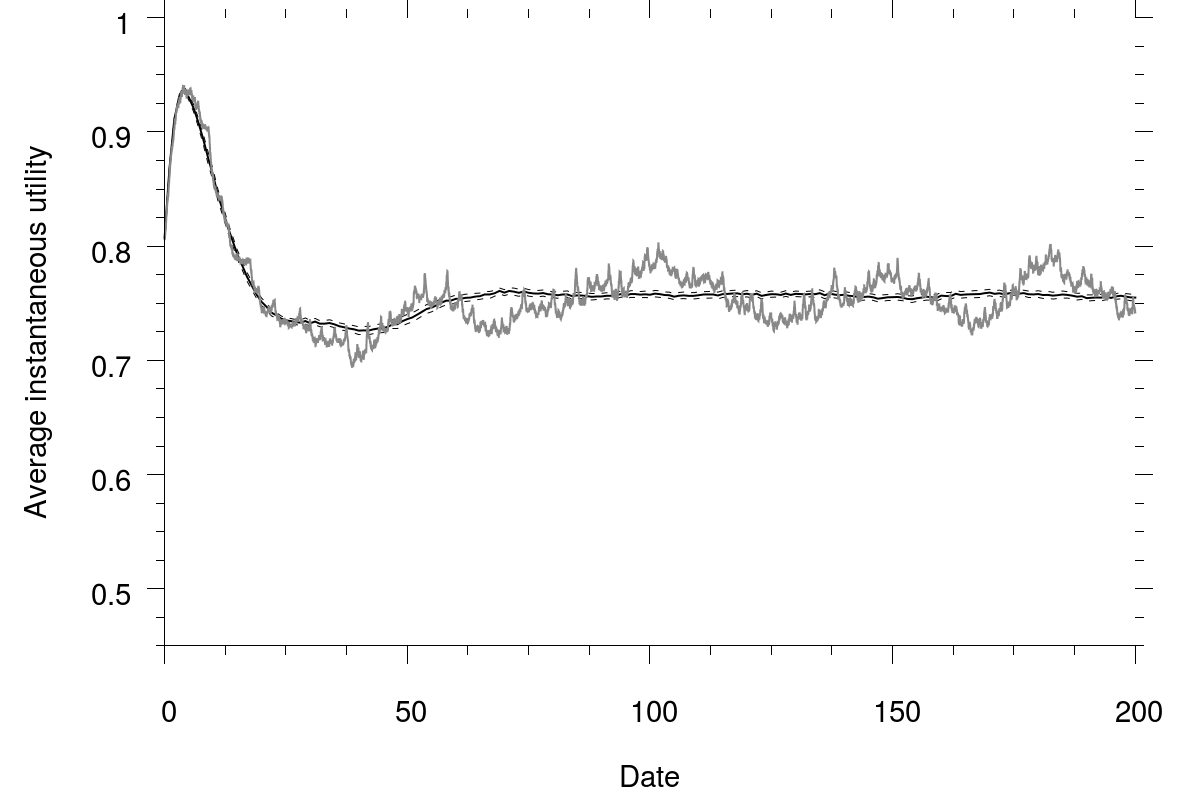}
\end{subfigure}%
\begin{subfigure}{.45\textwidth}
  \centering
  \includegraphics[width=.9\linewidth]{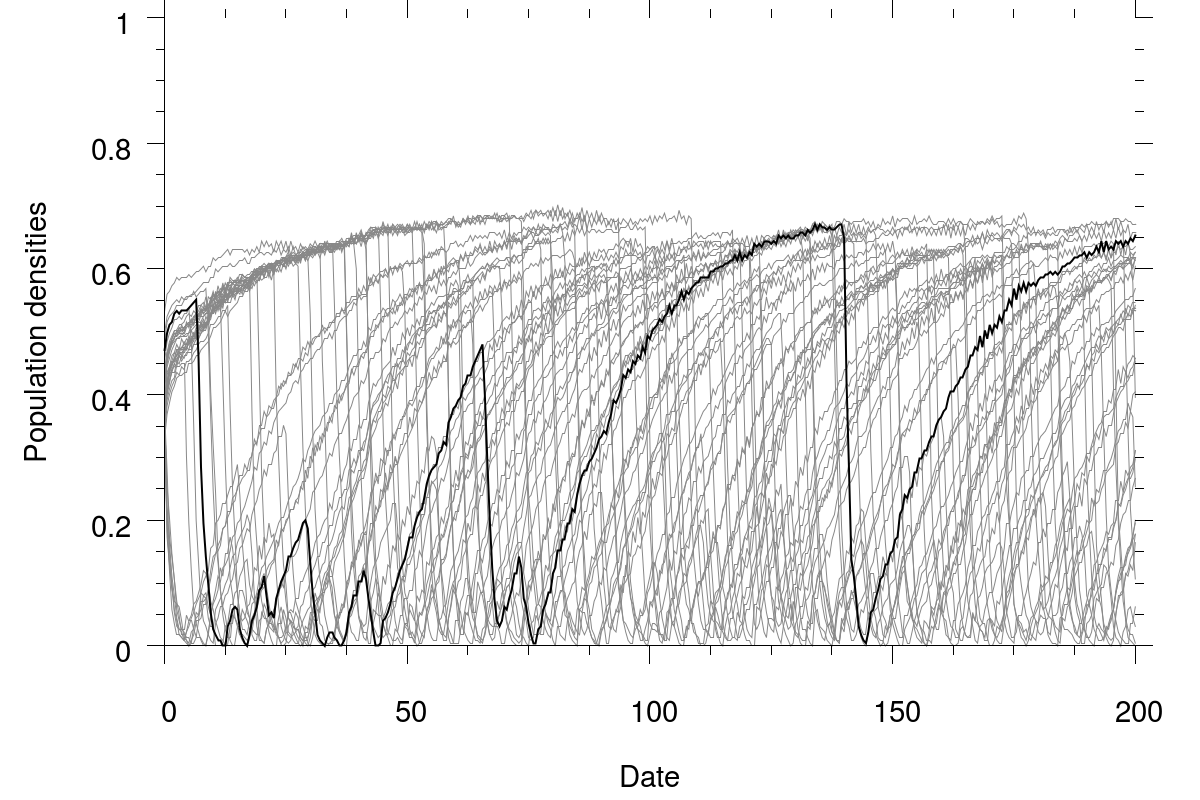}
\end{subfigure}
\caption{Model dynamics with $\underline{m}=1$ and $\overline{m}=10^{2.25}$. \textit{Left:} Average instantaneous utility with 95~\% confidence interval computed over 120 repetitions (one particular repetition in grey). \textit{Right:} Neighborhoods population densities for one repetition (in black one particular neighborhood).}
\label{fig5}
\end{figure}

Finally, when Step \ref{st3} does not hold, we can observe oscillatory dynamics. In Figure \ref{fig5}, we show such a dynamics obtained with $\overline{m}=10^{2.25} \approx 177.8$ and $\underline{m}=1$.\footnote{An alternative dynamics where Step \ref{st3} is not satisfied transitory is displayed in Figure \ref{fig1A} in Appendix.}

\section{Conclusion}\label{sec:conclu}

In this article, we introduced, in a framework very close to \cite{Jensen2018}, forecasting agents. These agents make their moving decisions using expectations for the future applying a linear extrapolation of the past. This formalization allows us to simulate a segregation model stochastically and yet have agents forming forecasts. We showed that for a relatively narrow set of parameters, this taking into account of expectations could solve the sub-optimality of decentralized decision making in segregation models.

In order to conclude, we would like to make a couple of remarks regarding the extrapolation heuristic we used in this article. First, we do not claim any uniqueness relatively to any desirable feature it may have. The only justification we have is its intuitive superiority over the fully myopic agents assumption implicitly set in all segregation models. Second, we are aware that some features of this heuristic are not desirable. In particular and importantly, notice that because the population densities and their forecasts are bounded between 0 and 1 for each neighborhood, there is no reason that individuals forecast densities that always have the same constant sum, $\frac{N}{Q.H}$ with our notation. Yet, forecasts with linear extrapolations seem one reasonable assumption to us even though many other heuristics could be set. The interest of our study rather lies in the originality of the study of forward-looking agents --~contrary to fully myopic until now~-- in segregation model. We sincerely hope that it will lead to more systematic studies in this direction.

\pagebreak

\bibliography{Schelling}

\pagebreak

%%%%%%%%%%%%%%%%%%%%%%%%%%%%%%%%%%%%%%%%%%%%%%%%%%%%%%%%%%%%%%%%%%
\clearpage
\appendix
\setcounter{figure}{0}
\renewcommand\thefigure{App-\arabic{figure}}    
\setcounter{table}{0}
\renewcommand\thetable{App-\arabic{table}}    
\setcounter{page}{1}
%%%%%%%%%%%%%%%%%%%%%%%%%%%%%%%%%%%%%%%%%%%%%%%%%%%%%%%%%%%%%%%%%%

\part*{Appendix}

\section{Additional results}

\begin{figure}[ht]
\centering
\begin{subfigure}{.45\textwidth}
  \centering
  \includegraphics[width=.9\linewidth]{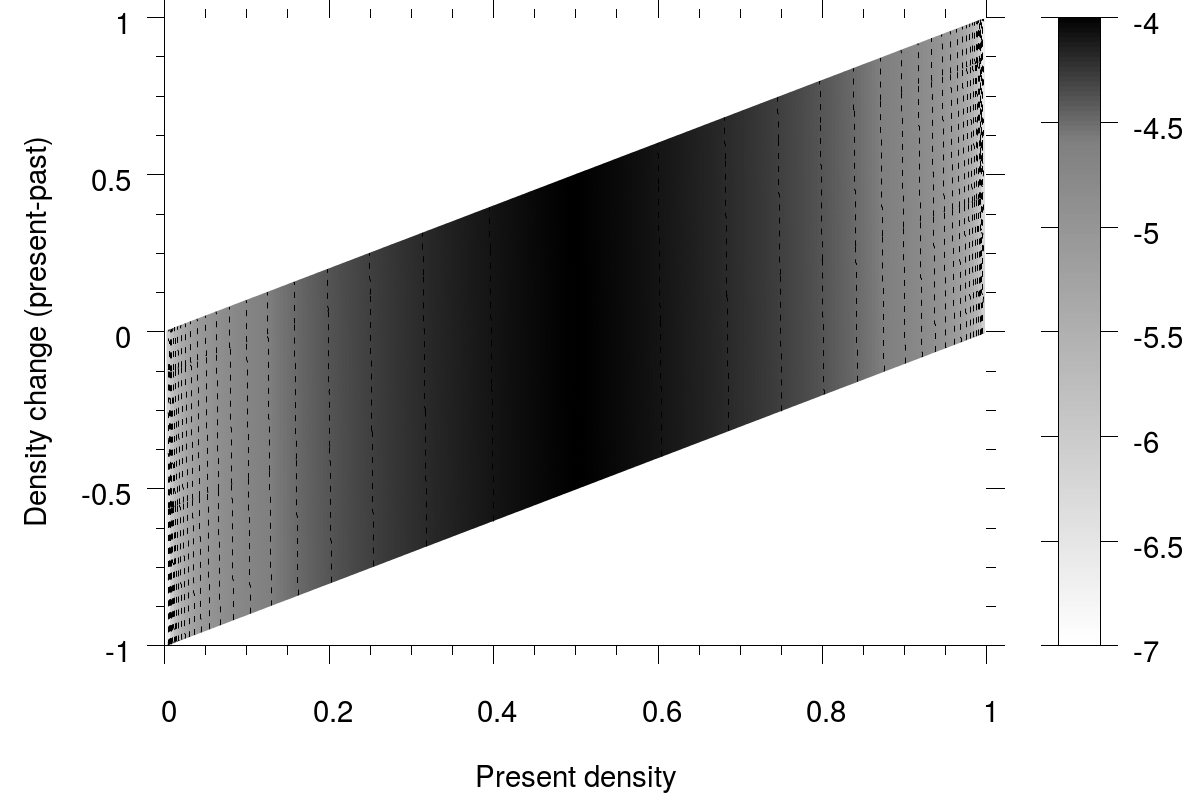}
\end{subfigure}%
\begin{subfigure}{.45\textwidth}
  \centering
  \includegraphics[width=.9\linewidth]{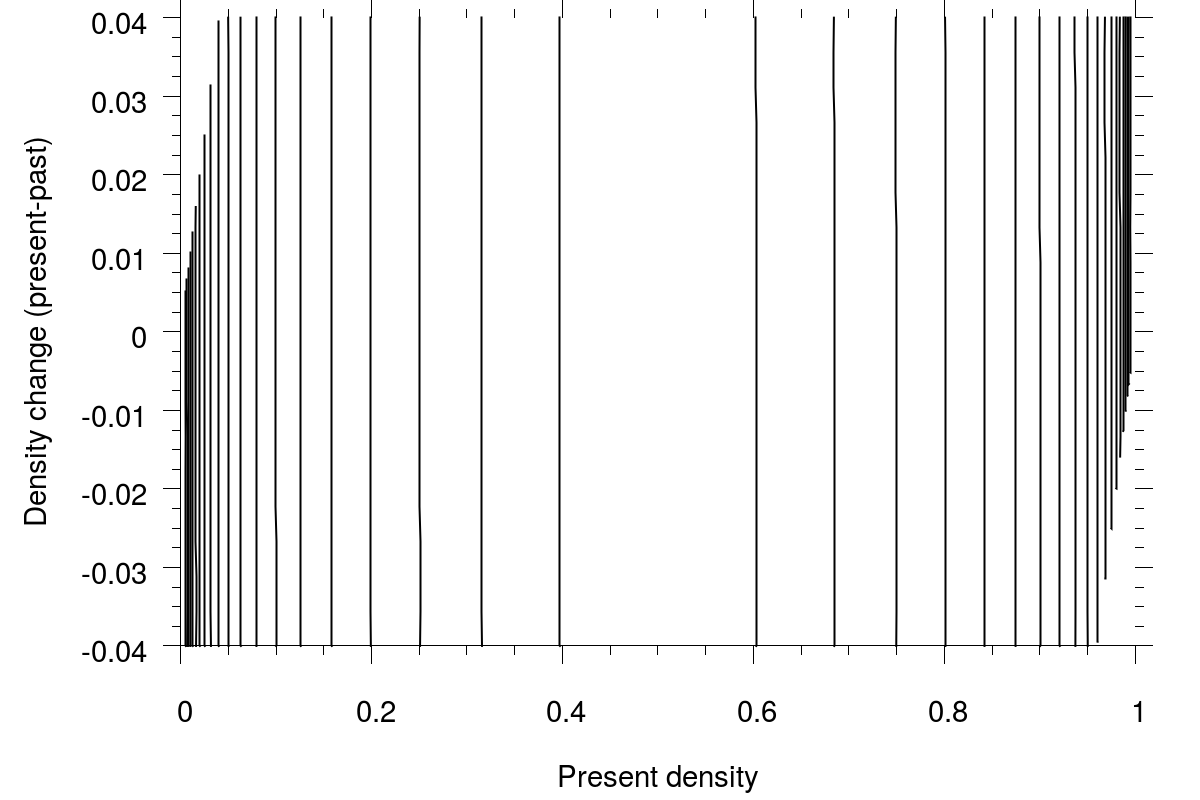}
\end{subfigure}
\caption{\textit{Left:} Expected intertemporal (computed over $\overline{m}=10^{-4}$) utility from living in a neighborhood as a function of the present population density in this neighborhood and the change in population density in this same neighborhood over the past $\underline{m}=10^{-2}$. Dashed lines for identical values contours. \textit{Right:} Same zoomed around 0 ordinate.}
\label{fig4A}
\end{figure}

\begin{figure}[ht]
\centering
\begin{subfigure}{.45\textwidth}
  \centering
  \includegraphics[width=.9\linewidth]{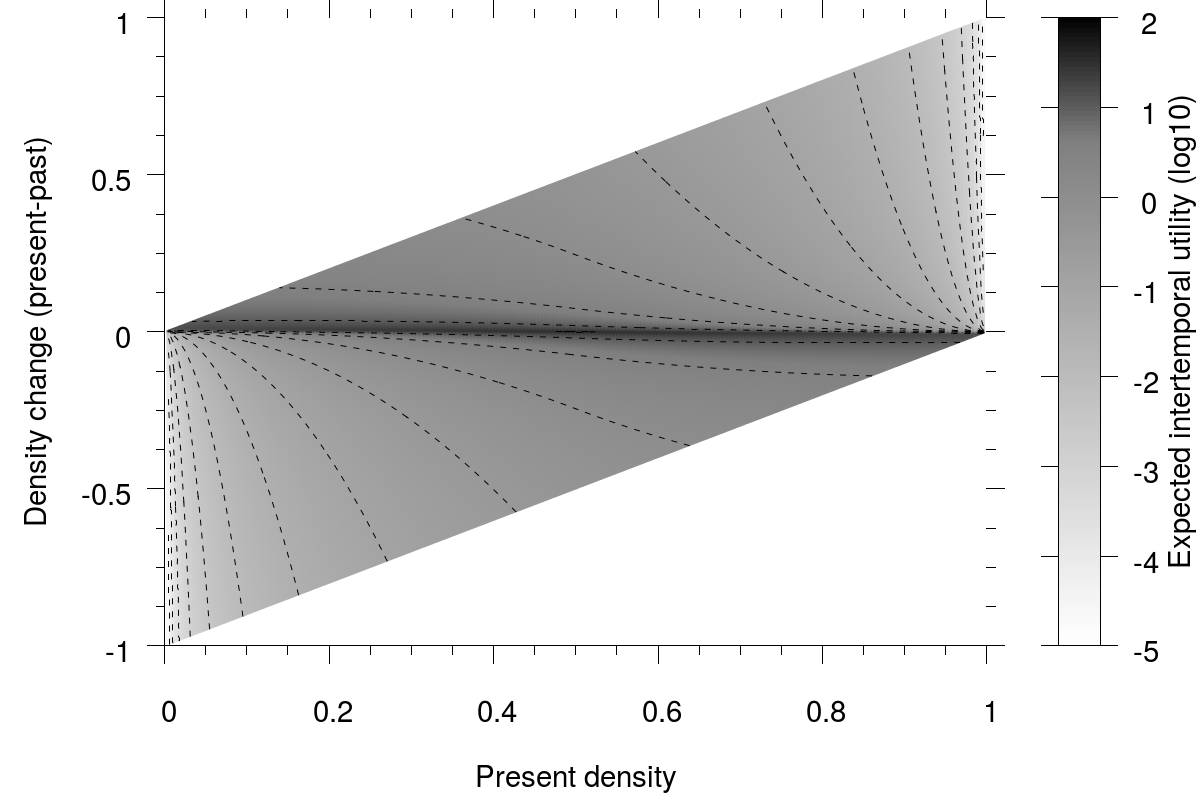}
\end{subfigure}%
\begin{subfigure}{.45\textwidth}
  \centering
  \includegraphics[width=.9\linewidth]{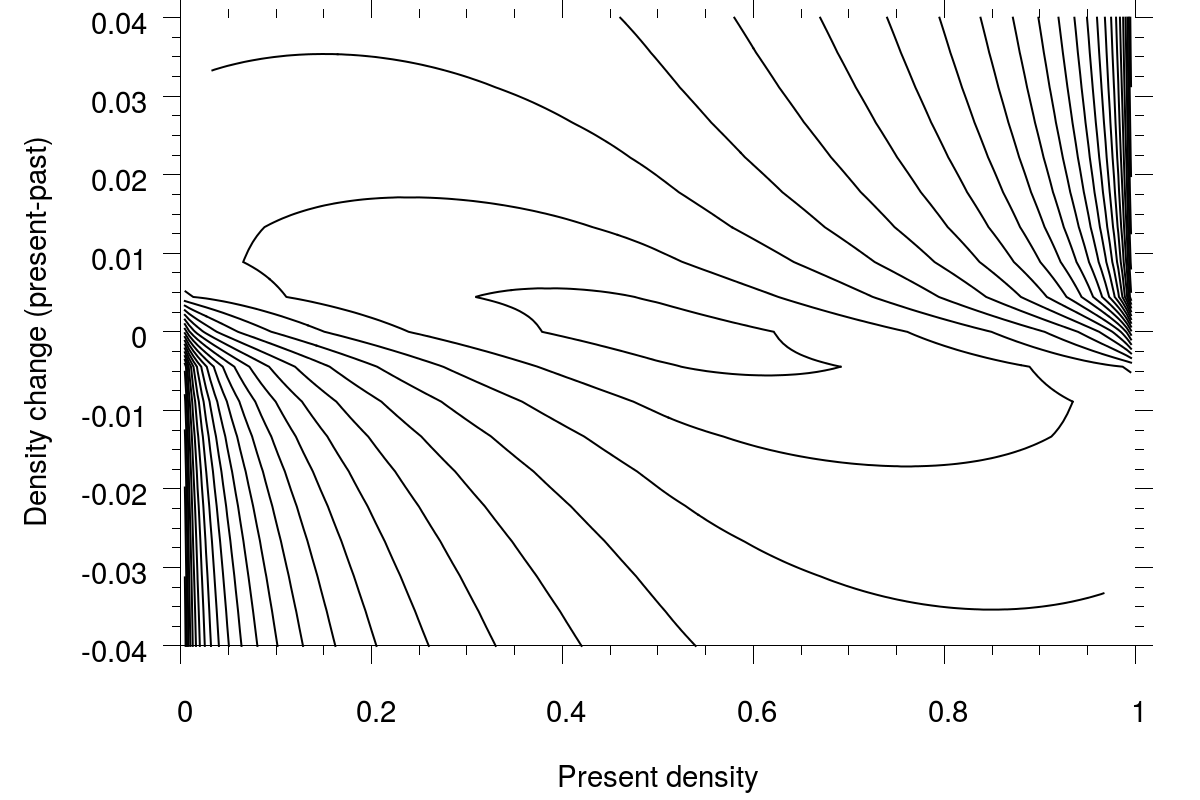}
\end{subfigure}
\caption{\textit{Left:} Expected intertemporal (computed over $\overline{m}=10^{2.25}$) utility from living in a neighborhood as a function of the present population density in this neighborhood and the change in population density in this same neighborhood over the past $\underline{m}=1$. Dashed lines for identical values contours. \textit{Right:} Same zoomed around 0 ordinate.}
\label{fig5A}
\end{figure}

In Figure \ref{fig4A} and \ref{fig5A}, we display the expected intertemporal utilities depending on both the present population density in a given neighborhood and the change in population density in this same neighborhood. $\underline{m}$ and $\overline{m}$ are chosen so that they match values in Figures \ref{fig1} and \ref{fig5} of the manuscript respectfully.

\begin{figure}[ht]
\centering
\includegraphics[width=.9\linewidth]{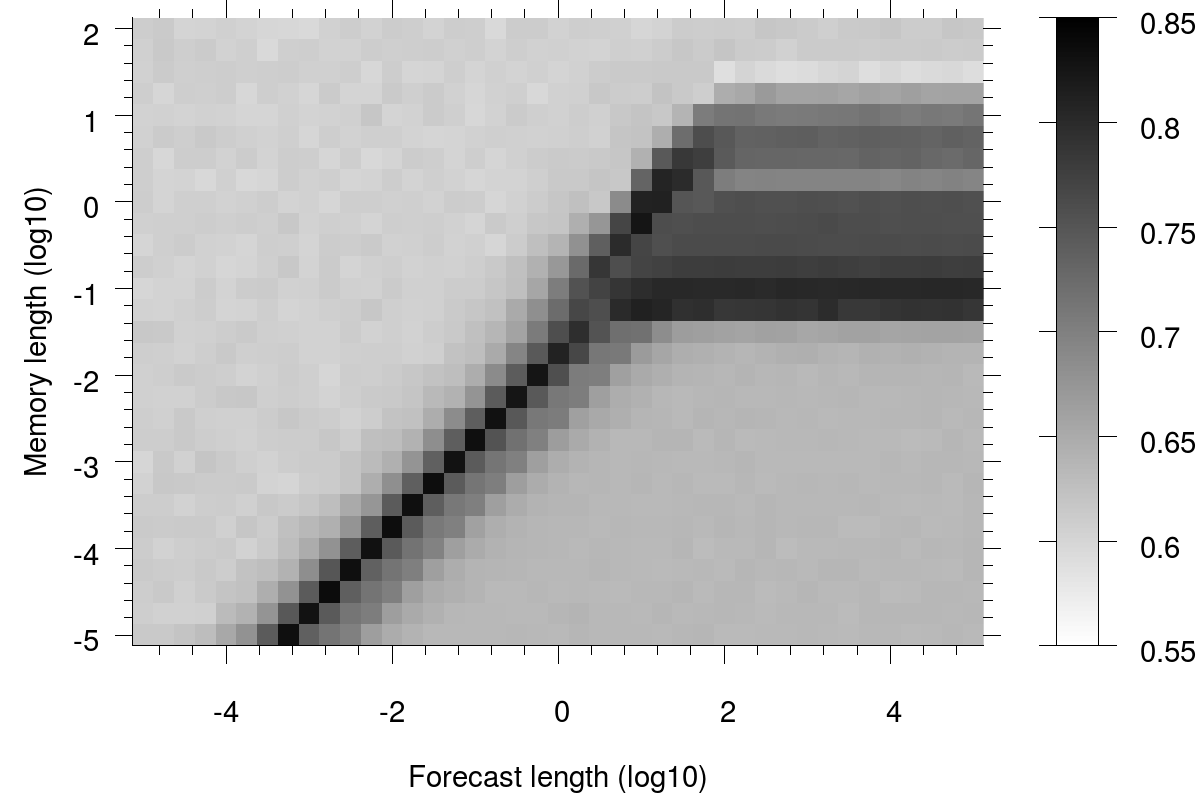}
\caption{Average instantaneous utility at stationary value as a function of the memory length $\underline{m}$ and the forecast length $\overline{m}$. Values averaged over 120 repetitions. Alternative initial conditions.}
\label{fig2A}
\end{figure}

In Figure \ref{fig2A}, we display the average instantaneous utility at its stationary value as a function of the memory length $\underline{m}$ and the forecast length $\overline{m}$ with initial conditions different from the ones described in the manuscript. Here, we consider agents and allocate them randomly in neighborhoods with probability $\frac{i}{\sum_{j=1,...,Q}j}$ for each neighborhood $i \in \{1,...,Q\}$ (1-indexed).

\begin{figure}[ht]
\centering
\begin{subfigure}{.45\textwidth}
  \centering
  \includegraphics[width=.9\linewidth]{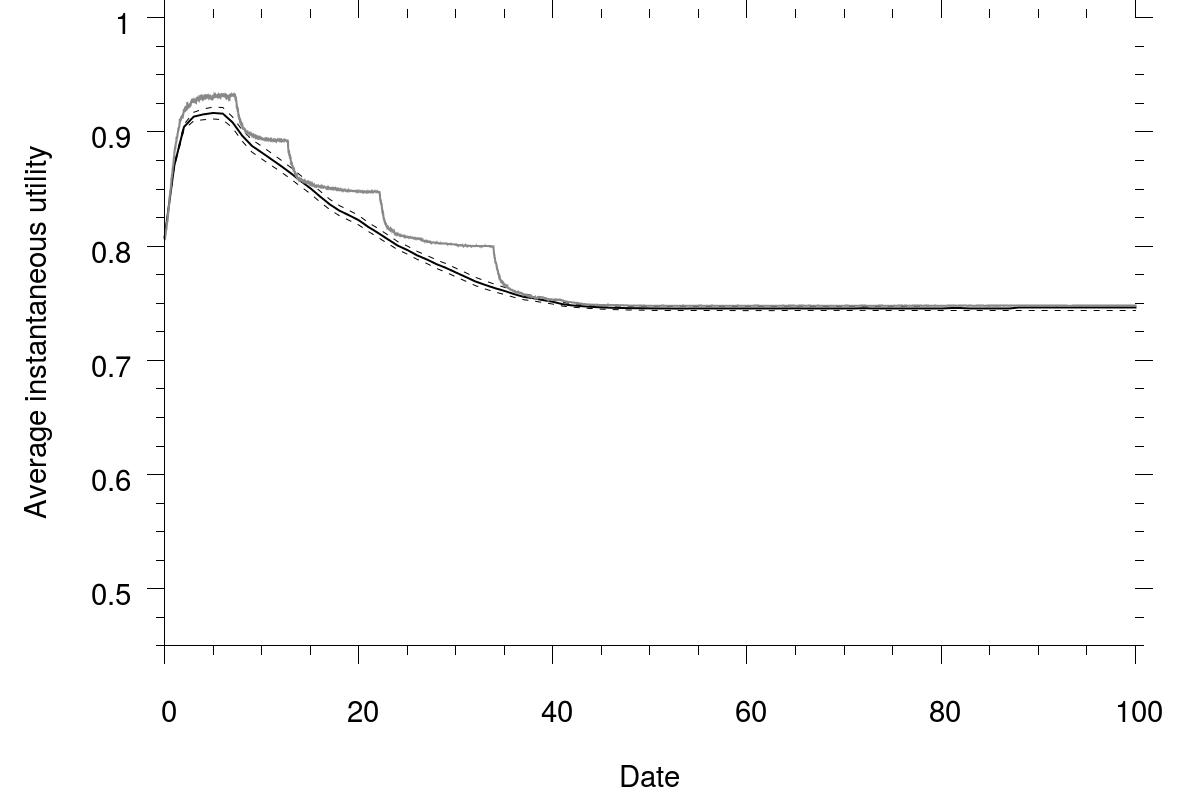}
\end{subfigure}%
\begin{subfigure}{.45\textwidth}
  \centering
  \includegraphics[width=.9\linewidth]{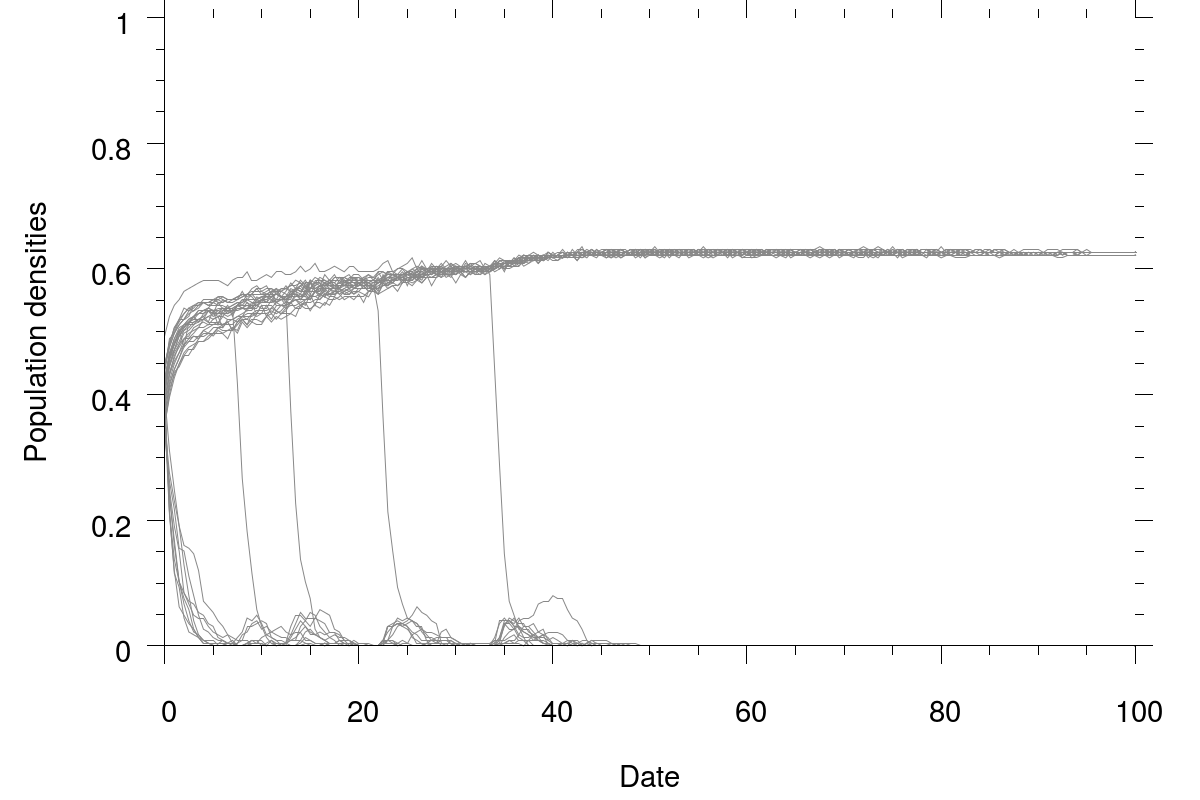}
\end{subfigure}
\caption{Model dynamics with $\underline{m}=10^{0.25}$ and $\overline{m}=10^{4}$. \textit{Left:} Average instantaneous utility with 95~\% confidence interval computed over 120 repetitions (one particular repetition in grey). \textit{Right:} Neighborhoods population densities for one repetition.}
\label{fig1A}
\end{figure}

\end{document}